# Effect of Clouds on Apertures of Space-based Air Fluorescence Detectors.

P. Sokolsky
High Energy Astrophysics Institute
University of Utah

J. Krizmanic
Universities Space Research Association
NASA Goddard Space Flight Center

**Abstract**

Space-based ultra-high-energy cosmic ray detectors observe fluorescence light from extensive air showers produced by these particles in the troposphere. Clouds can scatter and absorb this light and produce systematic errors in energy determination and spectrum normalization. We study the possibility of using IR remote sensing data from MODIS and GOES satellites to delimit clear areas of the atmosphere. The efficiency for detecting ultra-high-energy cosmic rays whose showers do not intersect clouds is determined for real, night-time cloud scenes. We use the MODIS SST cloud mask product to define clear pixels for cloud scenes along the equator and use the OWL Monte Carlo to generate showers in the cloud scenes. We find the efficiency for cloud-free showers with closest approach of three pixels to a cloudy pixel is 6.5%, exclusive of other factors. We conclude that defining a totally cloud-free aperture reduces the sensitivity of space-based fluorescence detectors to unacceptably small levels.

**Introduction**

There are several proposals to place Fly's Eye type air-fluorescence detectors in space. These include EUSO[1], under consideration for approval by the European Space Agency (ESA) for the International Space Station (ISS) at a 400 km orbit, and OWL[2], a proposal to put a pair of free-flying satellites in a higher (~1000 km) and near-equatorial orbit. These experiments would look down on the Earth's surface over latitudes ranging from near equatorial (+/- 5 deg. proposed for OWL) to the +/- 60 deg. accessible to the ISS. These detectors have wide-angle optics with half-opening angles of near 30 deg for EUSO and 22.5 deg for OWL. This corresponds to a footprint swept out over the Earth's surface by the near nadir pointing optical system of 170,000 and 540,000 km$^2$ respectively. Tilting the optical axis away from the nadir will increase the footprint area substantially, but we do not consider this possibility in this paper. The pixel size inside the footprint corresponds to 1km by 1km. Cosmic ray interactions in this footprint will be seen by the detectors through a broad range of weather conditions, mainly over the ocean's surface. The Fly's Eye technique[3] has been extensively developed by groups using upward-looking detectors placed on the Earth's surface. It utilizes the fact that ionizing particles in shower cascades, or extensive air showers (EAS) produced by incoming ultra-high-energy cosmic rays will excite $N_2$ fluorescence in the atmosphere. Detection of such fluorescence light (in the 300 to 400 nm UV region) can be used to reconstruct the shower energy and the shape and position of the cascade shower in the atmosphere can be employed to infer the composition of cosmic rays. Ground-based experiments have observed cosmic rays from ~ $10^{17}$ to just beyond $10^{20}$ eV. Predicted thresholds for OWL and EUSO range from $3 \times 10^{19}$ eV to near $10^{20}$ eV. The flux of cosmic rays above these energies is so low, that even such enormous apertures will yield only hundreds of events at the highest energies over the lifetime of the experiments.

*A critical issue for such space-based experiments is the fraction of aperture that is useful for the robust determination of cosmic ray shower energy and shower shape.* EAS produced by cosmic rays with energy greater than $3 \times 10^{19}$ eV will develop in the atmosphere and trigger the detectors as they traverse distances of between 10 and 100 km (depending on the zenith angle and the height of the initial interaction). Such showers may cross through and into cloud layer at various heights. In that case, the isotropically produced $N_2$ fluorescence light generated by the EAS will be multiply scattered in the cloud. In addition, the forward-going Cherenkov light beam which develops along the EAS will be effectively scattered by the cloud (both backscattered from the cloud top and multiply scattered in the volume of the cloud). Similar scattering will occur in aerosol layers, though these are mostly contained in the first few km of the atmosphere above the surface. Such scattered Cherenkov light will be picked up by the detector and produce very significant distortions superimposed on the shower profile produced by isotropic fluorescence light generated by the shower electrons.

Simulations have shown that passage of EAS showers through cloud layers will produce apparent structure in the shower profile[4]. In addition, since ~ 65% of observed showers at $10^{20}$ eV will have their shower maximum (Xmax) below 9 km above the ground[5], high cirrus clouds, occurring between 8 and 15 km will serve as an unpredictable attenuating mask. Light from the shower will pass through these clouds and be scattered. Unless the optical depth (OD) of these clouds as a function of position and UV wavelength is known, the energy and the shape of the shower will be mis-reconstructed.

Various techniques have been proposed to deal with this problem. The most promising is a LIDAR system mounted on the detector which would sweep a laser beam along the direction of the triggered event (passing through the same triggered pixels) within several seconds of the trigger. Back-scattered light would be detected either by the fly's eye detector itself, or by a specialized LIDAR receiver. This would detect the presence of even very thin clouds. Such information could be used to correct the signal, or veto the event as unreliable. A demonstration LIDAR system (Project LITE)[6] was flown on the Space Shuttle in 1995. While technical issues with the use of lasers in space are non-trivial, GLAS[7] (a laser-altimeter system) was launched in January 2002 with a planned three year operational life, and a planned two year duration satellite based LIDAR system (CALYPSO)[8] is set for launch in 2005.

It has been proposed that the intense Cherenkov beam in the shower can be used as an auto-diagnostic for the presence of clouds[1]. It is thought that cloud layers through which the shower passes will show up as structure superimposed on a smooth fluorescence light profile. More specifically, for optically thick and spatially thin clouds, the scattered Cherenkov light from the intense Cherenkov beam in the shower will develop peaks whose widths are related to the cloud thickness. However, optically thin but spatially thick (few km) clouds generate a much more subtle distortion of the shower shapes[4]. Showers passing through such clouds will be qualitatively similar to ordinary showers but may show unusually rapid rise or fall in their development. In the absence of clouds, such unusual showers would be a signature of new physics. Since such new physics would be of the greatest interest, such an auto-diagnostic technique precludes such discoveries. Unraveling the effects of clouds on shower development requires either a space-based LIDAR system to determine the locations of clouds along the triggered track (as is likely to be proposed for EUSO and OWL), or a stereo detector. In the case of a stereo detector such as OWL, the locations of peaks can be easily determined from stereo geometry alone. In addition, since scattered Cherenkov light has an angular dependence (dominated by the single-scattering phase function), the Cherenkov peaks or distortions will have different intensities when viewed at different angles by the two OWL detectors. In contrast, the portion of the shower which develops in clear air and is thus dominated by isotropically produced N2 fluorescence light will produce equal signals in the two stereo detectors after geometrical and atmospheric Rayleigh scattering correction. The lack of such balance for cloud scattered Cherenkov light will be an important signature, differentiating real from apparent "bumpy" structure in the development of an EAS in the atmosphere. In addition, the presence of high over-riding clouds that scatter the fluorescence light as it propagates towards the detectors will also manifest itself as an energy imbalance between the two stereo detectors.

*These techniques, while important for understanding signals produced by EAS, do not give the instantaneous aperture of the experiment*, i.e. what fraction of the geometrical aperture is sufficiently un-obscured to allow the EAS to trigger the detector. Since climatology studies indicate that clouds of one kind or other cover the Earth's surface about 70%[17] of the time, this is a non-trivial correction. This correction is also dynamic, constantly changing as a function of time as the detector footprint sweeps over the Earth's surface.

A LIDAR could in principle sample the entire aperture in a small enough grid to determine this aperture accurately. Unfortunately, the required data rate is much too high to be practicable, given the speed at which land passes below the orbiting detector. A coarse sampling is likely to be insufficient because cloud patterns and topologies are very variable on many distance scales. Furthermore, LIDAR, while pinpointing cloud locations accurately, does not represent how an EAS crossing the cloud would trigger the detector. This would have to be determined in Monte Carlo simulation, with some model of how this particular kind of cloud scatters light.

In this paper, we instead propose to ask a simpler question. Assuming that the details of the nature of clouds (height, OD, albedo etc.) are much more difficult to accurately ascertain than their simple presence, we inquire first into the fraction of the geometrical aperture which will be completely cloud free. If this is large enough, then the experiment can clearly be successful. If it is too small, then a next level of complexity must be addressed.

**IR Remote Sensing of Clouds**

We use existing remote-sensing data to develop cloud-masks. Since EUSO and OWL only operate at night, only IR data in the 3 to 15 micron region is useful (Note Solar reflected light represents a 6000 K black body, while nighttime IR from the Earth represents a ~ 270 K black body – hence IR above 3 microns will come from the Earth). There may also be differences in the kinds and distributions of clouds between day and night so we use only nighttime IR data. Combined GOES and other geostationary satellites give snapshots of the entire sub-polar Earth twice an hour[10]. However, the IR pixel size is not ideal (4km x 4km vs 1km x 1km for EUSO or OWL) and they have a limited number of wavelength windows. A new generation of GOES satellites (GIFTS) are planned for launch in 2006.[20] These will have imaging IR Fourier spectrometers, so that a full spectrum of IR light will be available for each pixel. This should make cloud height determination much more precise. At present, however, polar orbiting satellites that carry instruments such as MODIS have the best spatial and wavelength resolution (1km x 1km resolution and 36 spectral bands ranging from 0.4 to 14.4 microns). As discussed below, this information can be used to determine the height of the clouds more accurately.

**IR Transmission Through the Atmosphere and the SST**

IR is readily absorbed by water vapor and trace elements in the Earth's atmosphere at most wavelengths. However, there are a number of windows, notably 3-5 microns and 8-13 microns (see Fig. 1) that allow more efficient detection of IR from the Earth's surface.

MODIS and GOES satellites observe upwelling radiation near 11 microns, near the peak of the 270 K Black Body spectrum. In the absence of clouds and aerosols, the intensity of 11 micron IR is directly related to the surface temperature and surface emissivity. Ocean water is a good black-body, hence clear, cloud free pixels at 11 microns can be used to measure the sea surface temperature or SST. This "product" is of great interest to oceanographers and climatologists[11]. The SST, while having geographical and long-term temporal variations (cf. the "el nino" effect) is quite stable in the short-term. Its determination in a pixel can be verified using the extensive sea-buoy and freighter data base maintained by NOAA. What is required is knowledge that the pixel under consideration is truly "cloud-free".

The MODIS group has developed a set of algorithms (described below) to determine such cloud-free pixels[12]. They produce a "MODIS cloud-mask product" with four confidence indexes (high confidence cloud free, confident cloud free, probably cloudy and cloudy). Note that the SST cloud mask does not differentiate between high and low clouds or the presence of aerosols detected as clouds. This product has been checked by comparing with the SST derived from surface measurements.

A good example of this is the SST product for the Gulf of Mexico. It turns out that the Gulf of Mexico is essentially a perfect isotherm from the months of June to September[13] (less than 0.1 K SST variation). This makes it a perfect background for checking the cloud mask, since even thin clouds will produce a lower effective SST temperature relative to the uniform cloud-free pixels. Below, we use the MODIS cloud mask to determine the OWL cloud-free aperture efficiency.

**Cloud Detection Algorithms**

A number of algorithms have been developed to select out cloudy pixels[14]. They are based on four basic ideas:
      a. Temperature threshold
      b. Spatial coherence
      c. Temporal coherence
      d. Temperature differences for adjacent IR bands
We consider these in turn.

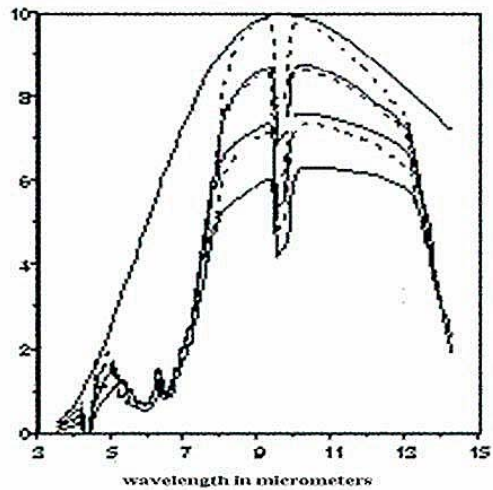

**Fig. 1 Earth radiance in the mid-to far-infrared spectrum. The various curves give a range of expected infrared radiances for a variety of typical atmospheres and surface temperatures. A 300 K blackbody curve is provided to permit visual comparison of path length absorption (from reference 18).**

### a. Temperature threshold

As indicated above, sea-water temperature is stable, with small diurnal variation and extensive surface data bases exist. Cloudy pixels will produce a lower temperature, with higher clouds appearing cooler than lower clouds (the lapse-rate of the atmosphere is approximately 6 degrees K/km). One can establish a threshold temperature, typically for the 11 micron IR window, $T_b$, such that pixels with $T < T_b$ are considered cloud contaminated. This threshold can be dynamically adjusted, either by comparing to the local geographical data base, or using the fact that any large enough cloud scene will have enough clear pixels that these will show up as a high temperature peak in a histogram (see Fig. 2).

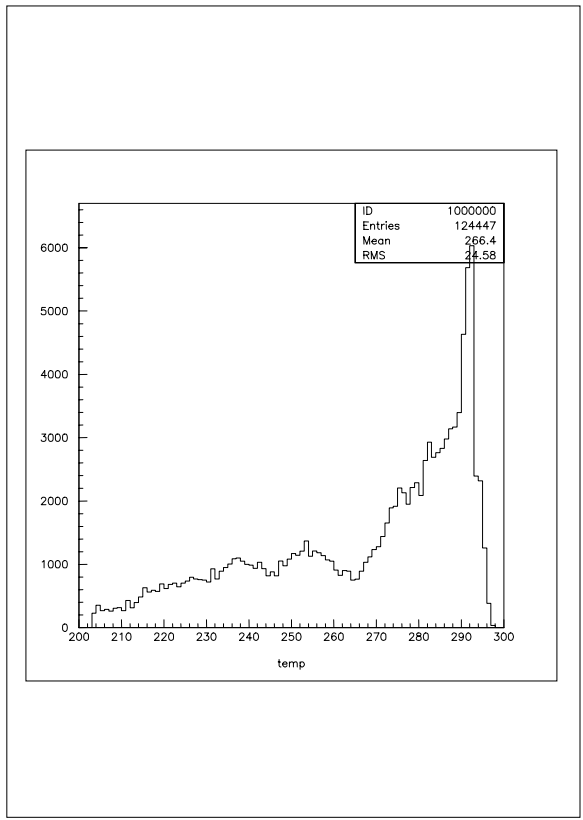

**Fig. 2. Typical Distribution of 11 micron IR radiance temperatures for GOES data over Equatorial Ocean for OWL footprint. Plot shows pronounced clear-pixel peak for T>292 K, consistent with the surface ocean temperature. The lower temperature distribution reflects the incidence of clouds at various altitudes. Higher altitude clouds have lower temperatures.**

### b. Spatial Coherence

Clouds often have large variations in effective IR temperature over small distance scales, due to changes in height and emissivity. An array of pixels (arrays as small as 3 x 3 are effective) can be used to determine a mean T and a standard deviation σ. If σ is larger than some threshold $\sigma_{thr}$ determined from arrays of clear pixels (as defined by the threshold temperature test, for example), then the entire array is flagged as potentially cloudy.

### c. Temporal Coherence

GOES and other geostationary satellites can check the stability of a pixel temperature as a function of time. Over water, clear pixels will show only the small diurnal variations.

Polar orbiting MODIS satellites typically return to the same scene about 9 times a day and similar criteria can be applied, but over considerably larger time intervals.

**d. Temperature Differences**

Radiances in adjacent IR windows in the 11-14 micron region come from different altitudes in the troposphere due to increasing $CO_2$ absorption with increasing wavelength. The 11 micron window sees surface radiances clearly while windows near 14 microns are only sensitive to IR from high cloud tops near 10 km, since radiation from below is absorbed. Distributions of temperature differences between such windows, $\Delta T_{ij}$, can be studied to establish a "clear" range and threshold rejection can be performed. Alternatively these temperature differences can be used in the study of spatial and temporal coherence. *This $\Delta T_{ij}$ test is particularly sensitive to the presence of high thin clouds which can be missed in a simple temperature threshold test.*

All of these tests can be combined to generate a cloud mask such as the MODIS SST product.

**Determination of Cloud Height**

Single, optically-thick clouds are assumed to be in thermal equilibrium with the surrounding atmosphere. The 6 deg/km temperature lapse rate in the troposphere (or more precisely, a measurement of the P(T) profile using radiosonde data) could then allow us to determine the cloud-top height from a single measurement of the 11 micron IR radiance, if the clouds emissivity were known and under the assumption that all radiation was emitted at the cloud-top surface. Unfortunately, cloud emissivities vary depending on cloud composition (ice versus water droplets, for example) and even with ice crystal structure. An alternative method called $CO_2$ slicing[14] has been developed to deal with this problem.

**$CO_2$ Slicing Algorithm**

The MODIS team has developed an algorithm for cloud height determination based on the following assumptions[15]

a. The cloud-top emissivity is a slow function of wavelength in the IR
b. A detectable cloud (typically with optical depth of > .1) can be represented by radiation from the cloud top only (this is necessary to make the mathematical analysis tractable).

Above 11 microns, CO2 absorption reduces IR throughput from the surface. Taking ratios of IR measurements in adjacent windows in this wavelength range both removes the dependence on emissivity and increases sensitivity to cloud-top height. The MODIS team states that a combination of this technique and the temperature difference technique allows them to resolve cloud heights even when overriding thin cirrus clouds are present. High thin cirrus clouds are stated to be detectable down to OD of ~0.1. Note, however, that since UV light is scattered by clouds more effectively than IR[21], that this corresponds to a near UV OD threshold more like 0.2 to 0.3. The physical and mathematical basis for the CO2 slicing technique is presented in Appendix A.

**Large-scale Cloud Distribution**

An overall view of the problem posed by clouds can be had by examining the HRES data set (this was an imaging IR satellite preceding the MODIS era) averaged over 2 deg by 3 deg latitude-longitude bins over latitude ranges up to +/- 60 degrees[17]. We use data averaged over 6 years for the months of February and July (representing possible seasonal variations) and broken down into low (< 2km), medium (2 to 8 km) and high (8 to 17 km) cloud incidence, as determined by the $CO_2$ slicing algorithm. Note that if multiple clouds are present, the data reports the highest cloud height. Fig 3 shows the incidence of various types of clouds as a function of latitude for the two seasons for latitude between + 20 and – 20 degrees. Table 1 summarizes the data for latitudes between +60 and – 60 degrees. Several general trends emerge.

a. Low clouds are present at all latitudes at the 40-50% level.
b. Medium high clouds occur independently of latitude with an incidence of about 20% and then rise to 25% at high latitudes. High clouds are somewhat more prevalent near the equator, but the incidence declines very slowly and remains at the 12 to 15% level for all orbital inclinations.
c. Seasonal variations, integrated over the orbital paths are small.

**Table 1 – Incidence (in %) of Different Cloud Types for Different Orbital Inclinations in a 2 deg x 3 deg Pixel.** Averages over six years of data for month of Feb. Numbers in parentheses are for month of July. +/- numbers indicate standard deviation.

| Orbital Inclination | Low | Medium | High Cloud Incidence |
|---|---|---|---|
| 10 deg. | 49 +/-17 | 19+/-10 | 15+/-7 |
|  | (51+/-14) | (20+/-10) | (13+/-8) |
| 20 deg. | 45+/-17 | 17+/-10 | 14+/-8 |
|  | (48+/-17) | (18+/-10) | (12+/-9) |
| 40 deg. | 39+/-17 | 20+/-11 | 14+/-7 |
|  | (41+/-18) | (20+/-11) | (13+/-9) |
| 60 deg. | 39+/-17 | 25+/-12 | 12+/-7 |
|  | (39+/-18) | (23+/-11) | (12+/-8) |

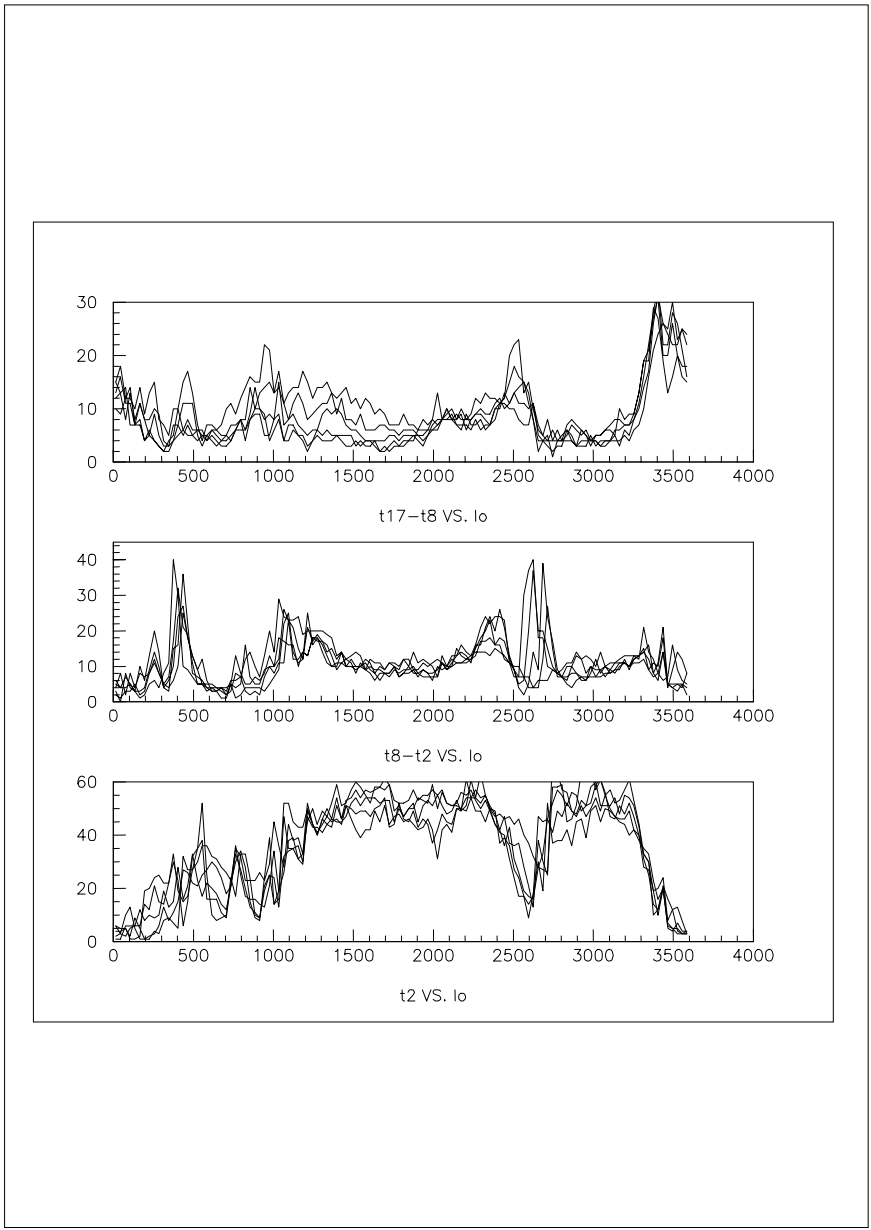

**Fig. 3a – Percent Incidence of Clouds (top – high(>8km), middle – medium (2-8 km), bottom – low (<2km) for 20 to 10 deg N latitudes (each trace is a 2 degree step in latitude). X axis is longitude in units of 0.1 degree.**

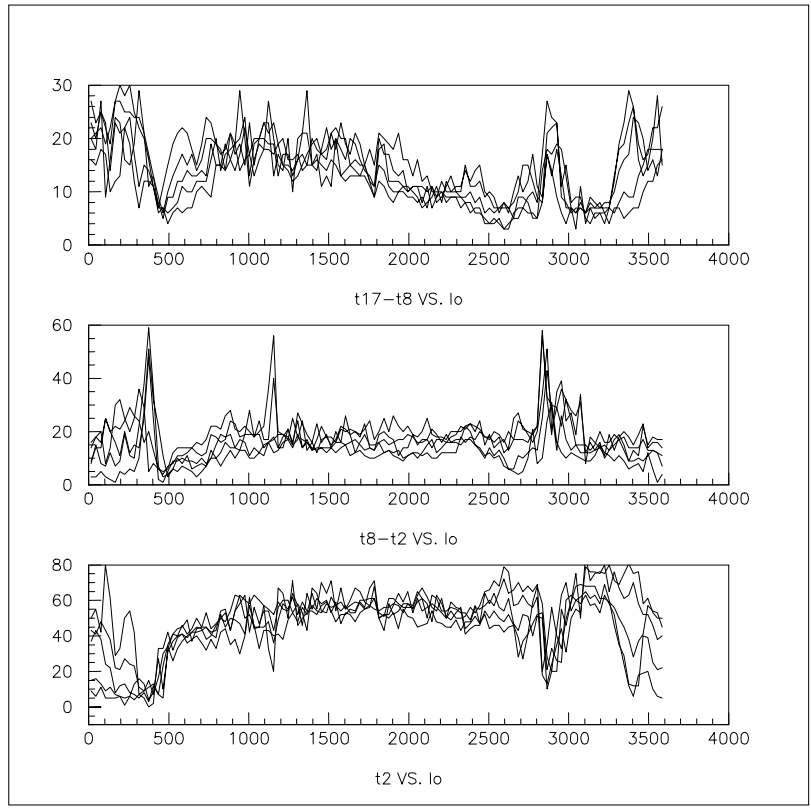

**Fig. 3b – Percent Incidence of Clouds (top – high(>8km), middle – medium (2-8 km), bottom – low (<2km) for 10 to 0 deg N latitudes. X axis is longitude in units of 0.1 degree.**

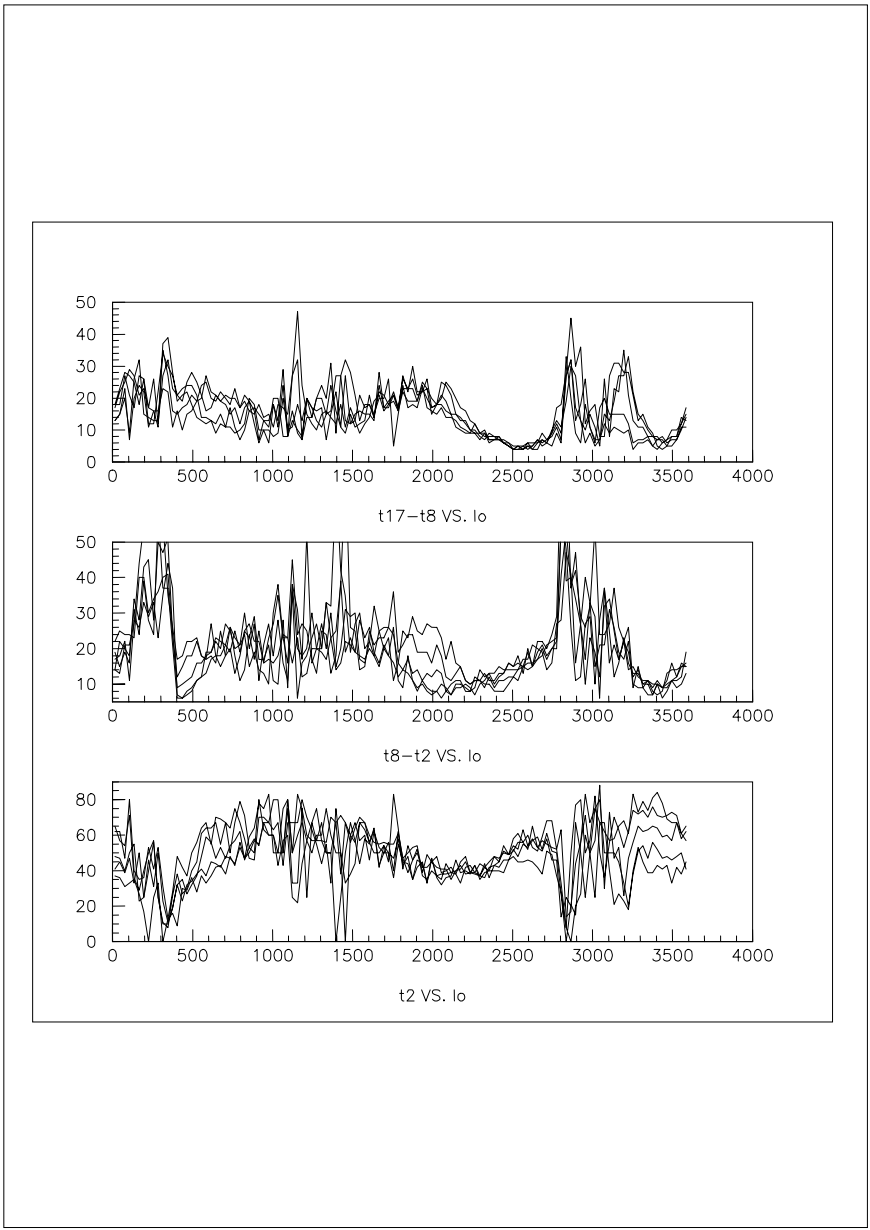

**Fig. 3c – Percent Incidence of Clouds (top – high(>8km), middle – medium (2-8 km), bottom – low (<2km) for 0 to -10 deg S latitudes. X axis is longitude in units of 0.1 degree.**

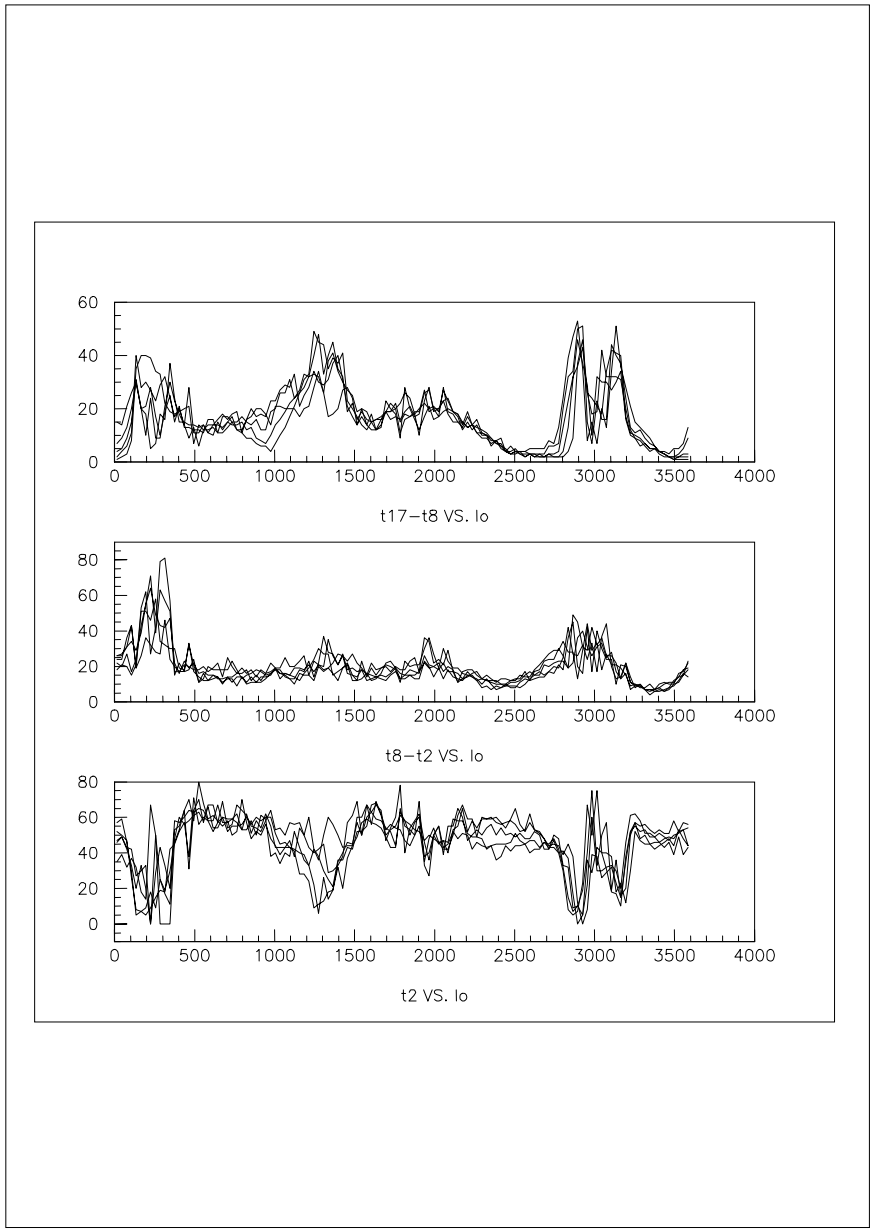

**Fig. 3d – Percent Incidence of Clouds (top – high(>8km), middle – medium (2-8 km), bottom – low (<2km) for -10 to -20 deg S latitudes. X axis is longitude in units of .1 degree.**

While there are areas on the Earth's surface which are relatively free of clouds (such as the South Pacific), integrated over all longitudes, the latitude dependence of cloud incidence is quite slow. The ISS orbit might have somewhat fewer high and low clouds and somewhat more mid-level clouds on average. Mid-level clouds certainly are the most problematic as they occur where most of the EAS develop, but the fraction of data taken over land (where cloud finding is much more difficult and the CO2 slicing method less reliable) and over light-polluted areas is also an issue. The only significant way to decrease cloud incidence is to go into geostationary orbit over an area like the South Pacific. This is not practical at the present level of technology, since it would require ~ 100 m diameter optical apertures.

**High Resolution Cloud Distributions**

While the 2 deg x 3 deg averaged HRES data is useful to give a general picture of the problem, it neglects the effect of correlations between different cloud types and is too coarse to convolve with the CR track-length distribution so as to determine the trigger aperture. To investigate this we take the most reliable remote-sensing based definition of a cloud-free pixel (derived from MODIS satellite data[9]) and create realistic cloud masks. We then throw Monte Carlo cosmic ray events into this real scene and require that the resultant track not cross any cloud-contaminated pixels and have a clear area (road) around it. In the case of scenes with thick, continuous cloud layers, we expect the efficiency to be very close to the ratio of clear to total pixels. For highly striated, chaotic or spottily dispersed clouds the efficiency depends on the topology, the fill factor and the length of track.

Specifically, to study the interaction of the track length distribution with the scale of clear spaces between clouds, we use the 1km x 1km MODIS SST cloud mask product from actual instantaneous cloud scenes[18]. These nighttime scenes are approximately 2200 × 2000 km$^2$ and are much larger than the ~ 400 km radius OWL footprint. We take the center of each scene and generate Monte Carlo events randomly throughout a footprint (see Fig 4 for a typical distribution).

The MODIS cloud mask product employs algorithms that incorporate the various IR band measurements along with ancillary data, e.g. land/water maps, to determine four clear-sky confidence levels. Numerically, a 99% confidence that a $1 \times 1$ km$^2$ pixel is cloud-free is denoted as *high-confidence clear*, a 95% confidence is considered *clear*, a 66% cloud-free confidence is considered *probably cloudy*, and a pixel is considered *cloudy* if the cloud-free confidence is less than 66%. For this study, we form a binary cloud flag for a given pixel by assigning a *high-confidence clear* or a *clear* mask value as CLEAR and a pixel with a *probably cloudy* or a *cloudy* mask value as CLOUDY. Given that the *probably cloudy* designation corresponds to a cloud-free probability of 66%, the inclusion of this mask value under the CLOUDY flag could, in principal, artificially enhance the level of cloudiness in a MODIS scene. However, the fraction of pixels with a *probably cloudy* mask value in a particular MODIS scene is approximately 10% for the scenes considered in this study with the location of the *probably cloudy* pixels highly correlated to the edges of *cloudy* regions. Thus, the conservative assignment of these as CLOUDY conforms to goal of this study: the determination of the fraction of observed UHECR airshowers that occur in definitely cloud-free areas of the viewed atmosphere.

The Monte Carlo events used in this cloud study assumed $10^{20}$ eV protons as the primaries and were randomly distributed uniformly in position and isotropic in angular incidence. Fully fluctuated airshowers were generated in 1 μsec time steps with the subsequent air fluorescence and scattered Cherenkov light attenuated by the atmosphere in a wavelength-dependent fashion. In the Monte Carlo, we assume that the atmosphere is cloud free. The OWL instruments were modeled with the 2002 baseline design[2] assuming 1000 km orbits and 500 km satellite separation. Events were accepted if they passed the nominal trigger criteria of having at least 4 detector pixels with at an integral signal of at least 5 photo-electrons (in each pixel) for both OWL eyes. The event sample included 1674 events, and the track length was defined as the portion of the airshower viewed by both instruments. The resultant 3-dimensional track length distribution is asymmetric with a mean value of approximately 16 km and a most probable value near 8 km. The two dimensional, xy-projection of the track lengths yields an asymmetric distribution with a mean value of approximately 15 km and a most probable value near 5 km. The xy-projected distribution has a range from slightly more than 0 km, corresponding to nearly vertical events,to approximately 125 km in projected length. The xy-projected track lengths did not include the modification of projecting onto a sphere as this is a minor effect for the spot size of approximately 400 km radius considered in this study.

The xy-projected track lengths were then superimposed on a sample of MODIS generated data samples that provided a pixel-by-pixel cloud mask with approximately 1 km spatial resolution. The MODIS data was from the 15th day of the odd months (Jan, Mar, etc.) in 2001 and at least 12 different MODIS near-equatorial measurements from each date were incorporated into this study. The center of each MODIS scene was selected for the OWL track superposition as each data scene was larger than the approximate 800 km diameter OWL ground spot size. The nearest distance of the various MODIS cloud mask designations (confident cloud, probable cloud, confident clear, high-confident clear) for a MODIS pixel as compared to the projected OWL track was recorded. Thus, the fraction of tracks with a cloud some minimum distance away could be determined for each MODIS scene.

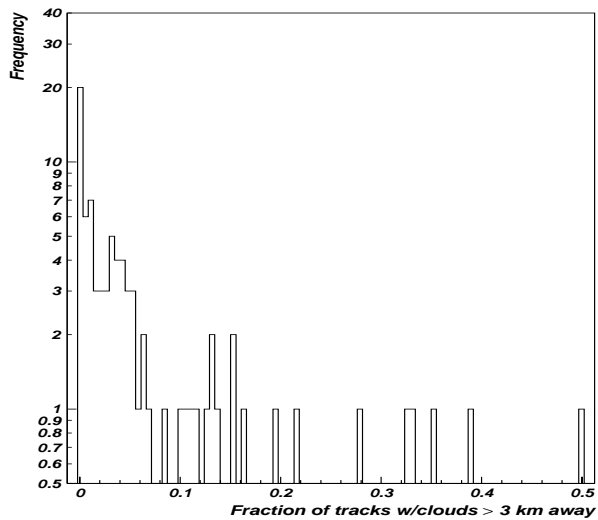

**Fig. 5 Distribution of Fractional Clear Aperture for 85 Randomly Selected MODIS Cloud Scenes Along the Equator. On average, only 6.5% of incident cosmic ray airshower tracks have a completely clear aperture.**

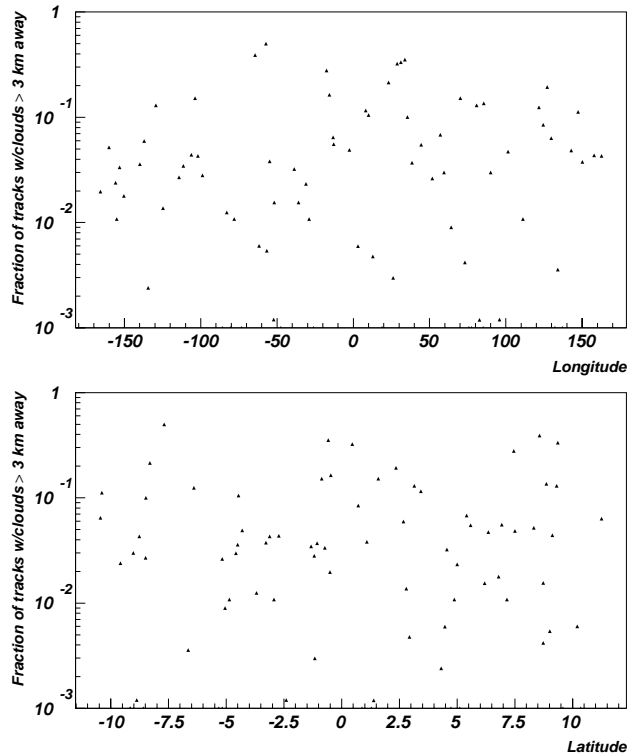

**Fig. 6 Fractional "Clear" Aperture as a Function of Longitude and Latitude.**

The least restrictive "clear aperture" is defined as having tracks with no clouds closer than 1 km or one pixel. Since the extended OWL optical spot size may split the signal between pixels, a more realistic aperture cut is defined for tracks with no clouds closer than 3 pixels. The more realistic "clear aperture" ranges from 0 to ~ 50% over the 85 randomly selected cloud scenes considered, *with the mean at 6.5%* and the median at 3.0 % (see Fig. 5). Fig 6 shows the distribution of the "clear" fraction as a function of latitude and longitude. There is no strong evidence for geographical correlation. Fig 7 shows the distribution of cloudy and clear pixels, and the correlation between the fraction of clear pixels and the fractional "clear" aperture for tracks.

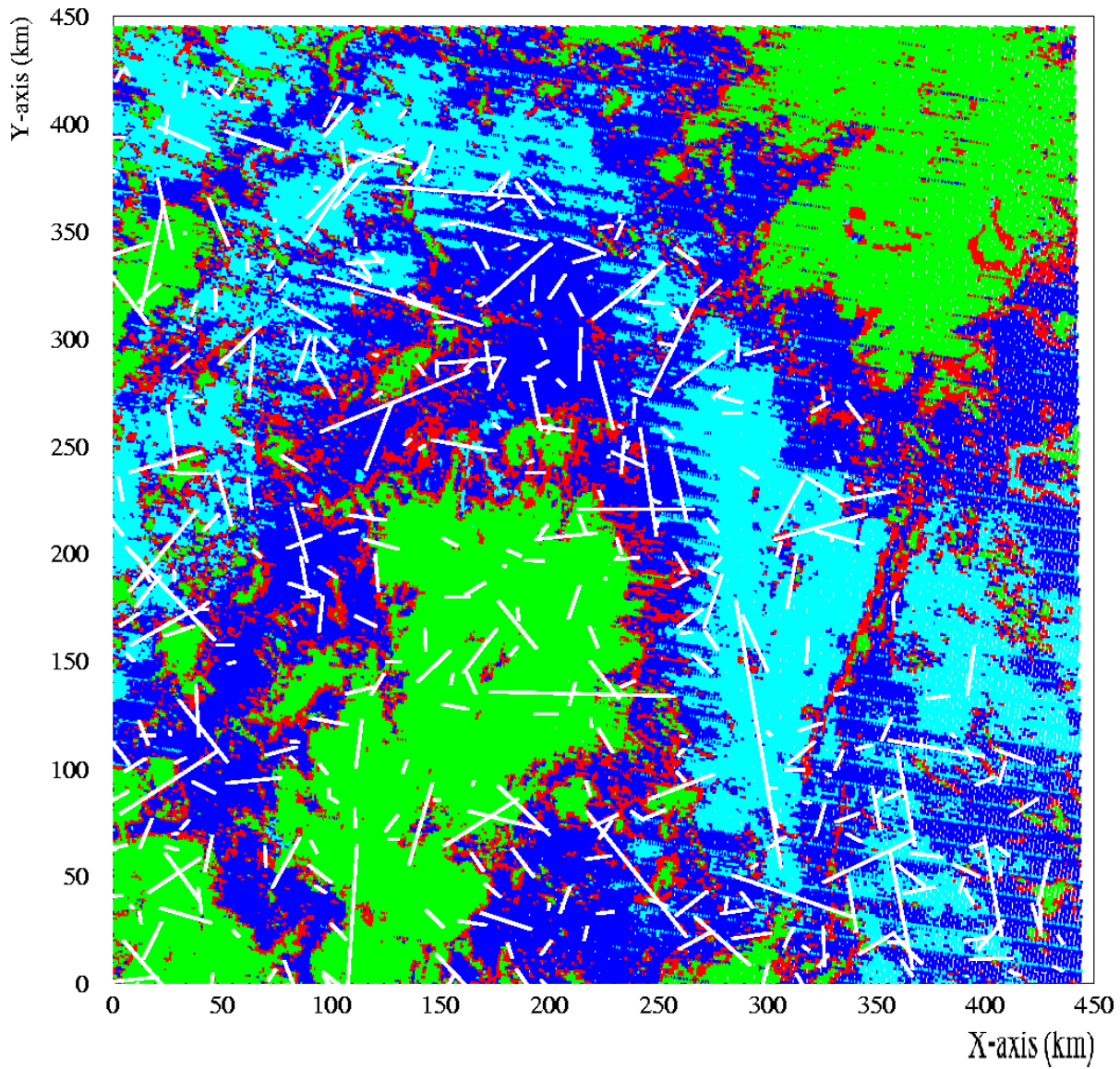

Fig 4. Portion of MODIS SST Cloud Mask. The projected $10^{20}$ eV simulated tracks, which trigger OWL, are shown superimposed as white lines. This view corresponds to approximately one quadrant of the square defined by the embedded quasi-circular OWL footprint. Light Blue – High-Confidence Cloud Free, Dark Blue – Cloud Free, Red – Probably Cloudy, Green – Cloudy. Note MODIS Roster Scanning Artifacts. For this particular Cloud Scene, only 19.9% of the simulated track sample have a "Clear" Aperture as defined by no cloudy or probably cloudy MODIS pixels within 3 km of a track.

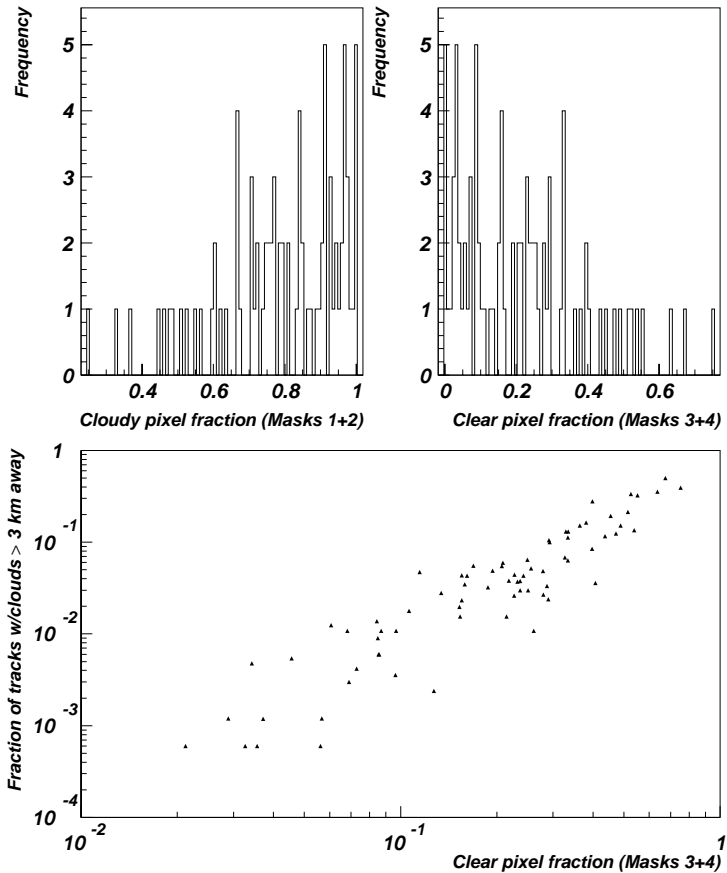

**Fig 7. Top: Distribution of Cloudy and Clear Pixels; Bottom: Correlation between fractional "clear" aperture and fraction of clear pixels.**

As expected from climatological studies, the overall cloudy pixel probability averages ~80% and the "clear" track aperture is significantly smaller than the clear pixel fraction, though it becomes approximately proportional for relatively clear cloud scenes. Preliminary results were based on half the number of cloud scenes reported here, but doubling the number did not significantly change the distributions.

This conservative estimate of "cloud-free" track efficiency results in a residual aperture of the full geometric aperture of only 6.5%. For the EUSO detector, its $1.7 \times 10^5$ km$^2$ footprint becomes effectively $1.1 \times 10^4$ km$^2$ which implies a time-averaged aperture (assuming a 14% on-time set by the requirement of no moon and no sun) of 1540 km$^2$, smaller than the Auger ground array area of 3000 km$^2$). For the OWL detector, its $5.4 \times 10^5$ km$^2$ footprint reduces to 4900 km$^2$ and while there will certainly be a significant number of "golden" cloud-free events near the OWL detector's low energy threshold, it is clearly necessary to deal with clouds in a less restrictive way to regain the lost aperture.

**Low Cloud Incidence**

Since low clouds (here defined as <2km) contribute about half of the total cloud incidence (see Table 1) and EAS typically have their shower maxima above 2km, we might expect a significant improvement in "clear" aperture if we are willing to live with such clouds. For monocular experiments, such as EUSO, their presence may even be helpful, since the UV albedo from low, dense water-vapor clouds is much higher than from sea or land. The reflected Cherenkov light from the cloud-top will produce a marker which can be used to improve the geometrical reconstruction of the track, *if the cloud height is known from some other measurement (LIDAR return, for instance)*. On the other hand, knowing that the clouds are indeed low and that there are no over-riding high thin clouds is a much less precise proposition than knowing that there are no clouds at all in a pixel.

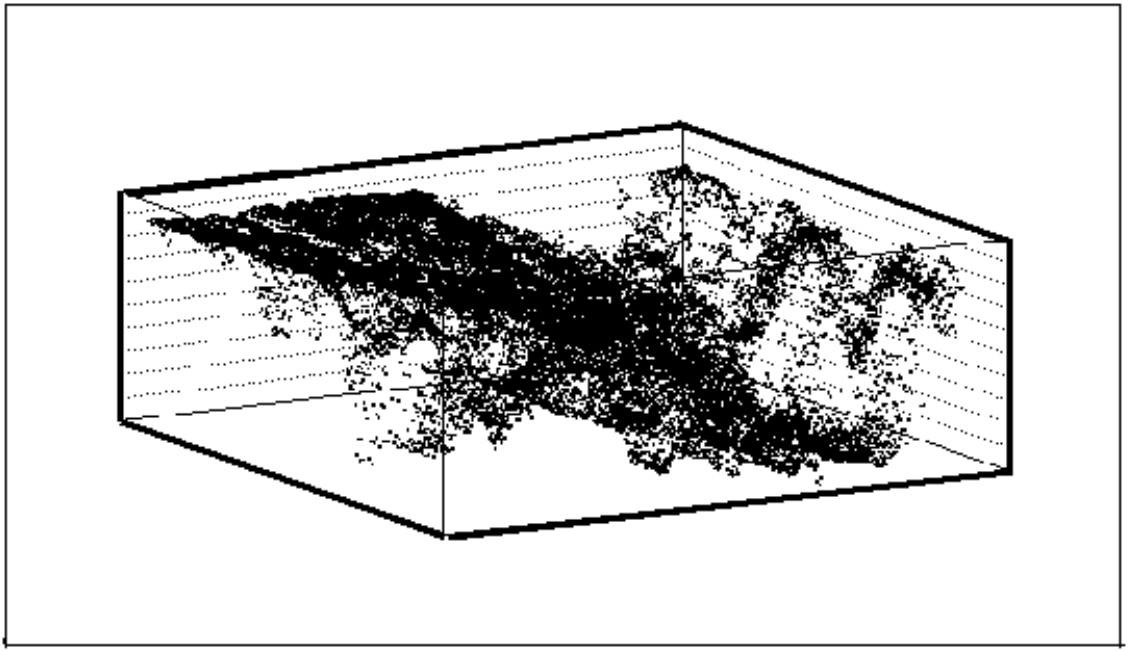

**Fig. 8a Distribution of GOES Pixel Temperatures for 800 km (latitude) and 1800 km (longitude) along the equator. Figure shows transition from low, warm clouds to high cool clouds back to low clouds again. Temperature is along Z axis, latitude along X axis, and longitude along Y axis.**

We are working with the MODIS team to produce a "clear or low-cloud" product, based on 11 to 13 micron temperature differences, the SST cloud mask, and the derived cloud-top temperature and pressure. For the purpose of getting a quick look, however, we use 11 micron GOES data[19] to find the effective cloud-top temperature for a particular day and hour. We crudely determine the SST from the distribution of warmest pixels in a cloud scene and assume a typical 6 deg/km lapse rate. For the data set under consideration, clear pixel sea surface temperature corresponds to T> 292 K, so that a 2 km cloud would have a T=280 K, while clouds near 10 km would have T=230 K or cooler (see Fig. 2). Figs. 8a and 8b show the distribution of cloud-top temperatures as determined in each 4 km by 4 km GOES 11 micron pixel as a function of latitude and longitude for a 800 by 1800 km swath near the equator. Fig 8a shows a transition from an area of low clouds to an area dominated by high clouds while Fig 8b shows an area with rapid variation between high and low clouds.

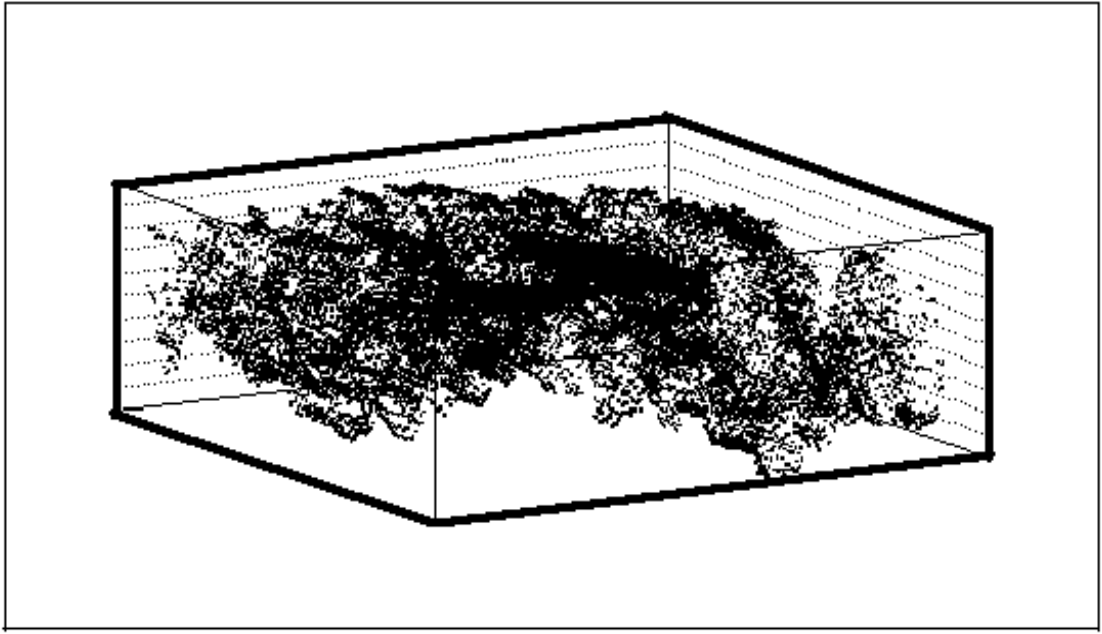

**Fig. 8b Distribution of GOES Pixel Temperatures for 800 km (latitude) and 1800 km (longitude) along the equator. Figure shows rapid transition from low warm clouds to high cool clouds on short distance scales. Temperature is along Z axis, latitude along X axis, and longitude along Y axis.**

We can also examine the distribution of pixel temperatures in each OWL aperture footprint around the equator. The ratio of clear (T>292 K) and clear or low cloud (T>280 K) to the total number of pixels for each footprint gives a crude upper limit on the "clear or low cloud" efficiency. For the particular day and hour considered, the clear efficiency was 31%, while the "clear or low cloud" efficiency was 70%, averaged over the equator. Over five contiguous 1800 km (longitude) x 800 km (latitude) steps, these efficiencies are 18 and 70%, 43 and 73%, 33 and 61%, 14 and 39% and 28 and 83% respectively. These efficiencies do not take into account the tracklength-cloud-topology interaction. While much more work is needed, particularly with the new and more reliable MODIS data, it seems likely that including low clouds will increase the useful geometrical aperture will increase by about a factor of two.

High thin cirrus clouds with optical depth of less than 0.3 may be difficult to detect over opaque low clouds. While such overlying clouds will certainly affect the energy and shower profile determination, the trigger efficiency is not affected[4], except very near detector threshold. Their presence can bias the determination of cloud-top pressure for low clouds to smaller values, however.

**Conclusions**

While space-based experiments have enormous geometrical apertures, the requirement of cloud-free viewing proposed in this paper imposes very stringent reductions. HRES data averaged over coarse bins indicates that ~75% of the time a cloud of some kind will be detected in a pixel. High spatial resolution MODIS data leads to a similar conclusion. The requirement of a cloud-free region of greater or equal to 3 km (3 pixels) around a track reduces the geometrical aperture by more than an order of magnitude to 6.5%. If complications due to the interaction of light from cosmic ray EAS with clouds are to be avoided, intrinsic geometrical apertures need to be very large. For example, assuming a 6.5% cloud-free fraction, and the EUSO estimate of 12% on time (34% darkness, 50% moon-free, 80% aurora-free) leads to an 0.8% overall efficiency. Because ground arrays can operate for a decade or more, while space-based detectors have typical lifetimes of three years, an additional factor of three is required to match the integrated exposure over the lifetime of the experiments. An order of magnitude larger aperture than the Auger ground detector (7000 km$^2$str) would thus require an intrinsic geometrical aperture of 2.7 x 10$^7$ km$^2$str.

We conclude that to fully exploit the space-based fluorescence technique, one must confront the issue of EAS – cloud interactions. This will be discussed in a paper presently in preparation.

**Appendix A – Basis of the $CO_2$ Slicing Method**

Consider a single level optically dense cloud in a single pixel. The cloud does not necessarily fill the pixel and the product of the filling fraction times emissivity is defined as $f$. In that case, the total upward-welling radiance at the MODIS detector, $R(\lambda)$ can be written as:

$R(\lambda) = R_{surface}(\lambda, T_s)(1-f) + fR_{cloud}(\lambda, T_c) + R_{below}(1-f) + R_{above}$

Where $R(\lambda, T)$ is proportional to the Planck function, $T_s$ and $T_c$ are the surface and cloud-top temperatures while $R_{below}$ and $R_{above}$ are the integrated column radiances from the atmosphere below and above the cloud top. Note in this approximation, atmospheric column radiances directly below the cloud are assumed to be totally absorbed by the cloud, and the cloud emission occurs at the cloud top only.

We can then define the clear radiance $R(\lambda)_{clear}$ as

$$R(\lambda)_{clear} = R_{surface}(\lambda, T_s) + R_{below} + R_{above}$$

Then $\Delta R = R(\lambda) - R(\lambda)_{clear} = -f R_{surface}(\lambda, T_s) + f R_{cloud}(\lambda, T_c) - f R_{below}$
where, more precisely,

$$R_{surface}^1(\lambda, T_s) = B(\lambda, T(P_s))\tau(\lambda, P_s)$$

$$R_{cloud}(\lambda, T_c) = B(\lambda, T(P_c))\tau(\lambda, P_c)$$

$$R_{below} = \int_{P_s}^{P_c} B(\lambda, T(P))(d\tau/dP) \, dP$$

where B is the Planck distribution function and $\tau(\lambda, P)$ is the atmospheric absorption from pressure P to the top of the atmosphere. T(P) is the temperature profile of the atmosphere.

Integrating by parts, one finds

$$\Delta R = f \int_{P_s}^{P_c} \tau(\lambda, P)(dB(\lambda, T(P))/dP) \, dP.$$

If observations are made at two windows with similar $\lambda$, then one can assume that $f$ is independent of wavelength and $\Delta R(\lambda_1)/\Delta R(\lambda_2)$ depends only on $P_c$ if $\tau(\lambda, P)$ and $T(P)$ is known. Now $\Delta R$ is a measured quantity since one can find a clear air pixel close to the cloudy pixel under consideration using the SST cloud mask. The RHS of the ratio equation can then be calculated and compared to the measured ratio $\Delta R(\lambda_1)/\Delta R(\lambda_2)$ for a series of nearby wavelengths. $P_c$ is then the best match for the whole series. Note that once $P_c$ is known, $f$ can be calculated as well.

## Aknowledgements

We would like to thank Bob Streitmatter for supporting this work and for many stimulating discussions. Bill Ridgeway and Dennis Chesters were generous with their time in pointing us to relevant MODIS and GOES data and reformatting data to fit our needs. One of us (P.S.) would like to thank the John Simon Guggenheim Foundation and Universities Space Research Association for financial support.